\documentclass[12 pt, amsfonts, amssymb,color]{article}

\usepackage{amsmath,amssymb,amsfonts,latexsym}

\begin{document}

\renewcommand\theequation{\arabic{section}.\arabic{equation}}
\catcode`@=11 \@addtoreset{equation}{section}
\newtheorem{axiom}{Definition}[section]
\newtheorem{theorem}{Theorem}[section]
\newtheorem{axiom2}{Example}[section]
\newtheorem{axiom3}{Lemma}[section]
\newtheorem{prop}{Proposition}[section]
\newtheorem{axiom5}{Corollary}[section]
\newcommand{\be}{\begin{equation}}
\newcommand{\ee}{\end{equation}}
\newcommand{\bel}{\begin{equation}\label}

\newcommand{\equal}{\!\!\!&=&\!\!\!}
\newcommand{\rd}{\partial}
\newcommand{\g}{\hat {\cal G}}
\newcommand{\bo}{\bigodot}
\newcommand{\res}{\mathop{\mbox{\rm res}}}
\newcommand{\diag}{\mathop{\mbox{\rm diag}}}
\newcommand{\Tr}{\mathop{\mbox{\rm Tr}}}
\newcommand{\const}{\mbox{\rm const.}\;}
\newcommand{\cA}{{\cal A}}
\newcommand{\bA}{{\bf A}}
\newcommand{\Abar}{{\bar{A}}}
\newcommand{\cAbar}{{\bar{\cA}}}
\newcommand{\bAbar}{{\bar{\bA}}}
\newcommand{\cB}{{\cal B}}
\newcommand{\bB}{{\bf B}}
\newcommand{\Bbar}{{\bar{B}}}
\newcommand{\cBbar}{{\bar{\cB}}}
\newcommand{\bBbar}{{\bar{\bB}}}
\newcommand{\bC}{{\bf C}}
\newcommand{\cbar}{{\bar{c}}}
\newcommand{\Cbar}{{\bar{C}}}
\newcommand{\Hbar}{{\bar{H}}}
\newcommand{\cL}{{\cal L}}
\newcommand{\bL}{{\bf L}}
\newcommand{\Lbar}{{\bar{L}}}
\newcommand{\cLbar}{{\bar{\cL}}}
\newcommand{\bLbar}{{\bar{\bL}}}
\newcommand{\cM}{{\cal M}}
\newcommand{\bM}{{\bf M}}
\newcommand{\Mbar}{{\bar{M}}}
\newcommand{\cMbar}{{\bar{\cM}}}
\newcommand{\bMbar}{{\bar{\bM}}}
\newcommand{\cP}{{\cal P}}
\newcommand{\cQ}{{\cal Q}}
\newcommand{\bU}{{\bf U}}
\newcommand{\bR}{{\bf R}}
\newcommand{\cW}{{\cal W}}
\newcommand{\bW}{{\bf W}}
\newcommand{\bZ}{{\bf Z}}
\newcommand{\Wbar}{{\bar{W}}}
\newcommand{\Xbar}{{\bar{X}}}
\newcommand{\cWbar}{{\bar{\cW}}}
\newcommand{\bWbar}{{\bar{\bW}}}
\newcommand{\abar}{{\bar{a}}}
\newcommand{\nbar}{{\bar{n}}}
\newcommand{\pbar}{{\bar{p}}}
\newcommand{\tbar}{{\bar{t}}}
\newcommand{\ubar}{{\bar{u}}}
\newcommand{\utilde}{\tilde{u}}
\newcommand{\vbar}{{\bar{v}}}
\newcommand{\wbar}{{\bar{w}}}
\newcommand{\phibar}{{\bar{\phi}}}
\newcommand{\Psibar}{{\bar{\Psi}}}
\newcommand{\bLambda}{{\bf \Lambda}}
\newcommand{\bDelta}{{\bf \Delta}}
\newcommand{\p}{\partial}
\newcommand{\om}{{\Omega \cal G}}
\newcommand{\ID}{{\mathbb{D}}}
\newcommand{\pr}{{\prime}}
\newcommand{\prr}{{\prime\prime}}
\newcommand{\prrr}{{\prime\prime\prime}}
\title{Superintegrable systems with position dependent mass:
master symmetry and action-angle methods \\}

\author {A. Ghose-Choudhury \footnote{ E-mail: {\tt
aghosechoudhury@gmail.com}}\\
Department of Physics, Surendranath  College,\\ 24/2 Mahatma
Gandhi Road, Calcutta-700009, India.\\
\and
Partha Guha\footnote{E-mail: {\tt partha@bose.res.in}}\\
S.N. Bose National Centre for Basic Sciences \\
JD Block, Sector III, Salt Lake \\ Kolkata - 700098,  India \\}

\date{}

\maketitle

\begin{abstract}
We consider the issue of deriving superintegrable systems with
position dependent mass (PDM) in two dimensions from certain known
superintegrable systems using the recently introduced method of
master symmetries and complex factorization by M. Ranada \cite{Rana1,Rana2,Rana3,Rana4}.
We introduce a noncanonical transformation
to map the Hamiltonian of the PDM systems to that of ordinary unit mass systems.
We observe a duality between these systems. We also study Tsiganov's method
\cite{Tsiganov1,Tsiganov2,Tsiganov3,Tsiganov4} to derive
polynomial integrals of motion using addition theorems for the action-angle variables
using famous Chebyshev's theorem on binomial differentials. We compare Tsiganov's  method of generating an 
additional integral of motion with that of Ranada's master symmetry method.

\end{abstract}


\smallskip

\paragraph{Mathematics Classification (2000)}: 37J15, 70H06.

\smallskip

\paragraph{Keywords and Key phrases :} Superintegrable systems;
Master symmetries; Complex factorization; Action-angle variables; Position dependent mass.

\section{Introduction}
It is well known that a Hamiltonian system with $n$ degrees of
freedom is integrable in the Liouville sense if it possesses $n$
functionally independent constants of motion which are in
involution.  If $H$ denotes the Hamiltonian then there must exists
constants of motion $I_a$ such that $$\{H, I_a\}=0,
\;\;\;a=1,2,...,n-1$$  with $$\{I_a, I_b\}=0,
\;\;a,b=1,2,...,n-1.$$  It is generally true that integrable
systems are exceptional as most dynamical systems governing
physical phenomena rarely possess the requisite  number of
constants of motion to ensure their integrability. Having said
this it is interesting to note that there are systems which
possess even more integrals of motion then that required by
integrability. Such systems are generally termed as
superintegrable. The formal definition of such systems may be
stated as follows.\\
An integrable system is superintegrable if it allows
additional integrals of motion $J_b(q,p)$ such that $\{H, J_b\}=0$
for all $b=1,2,...,k$ with the set $\{ H,
I_1,...I_{n-1},J_1,...J_k\}$ being functionally independent so
that
$$\mbox{rank}\frac{\partial(H,I_1,...,I_{n-1},J_1,...,J_k)}{\partial(q_1,...q_n,p_1,...,p_n)}=n+k$$

\smallskip

In \cite{Fris}  Fris {\it et al}   studied  systems  admitting
separability in two different coordinate systems in the Euclidean
space and obtained four families of potentials possessing three
functionally independent integrals of the motion. The first family
of such potentials,  is known as the Smorodinsky-Winternitz (SW)
potential and is a two-dimenensional generalization of the
isotonic oscillator. The superintegrability of the SW potential
has also been investigated by Evans \cite{Evans,VE} in the more
general case of $n$ degrees of freedom. In general integrable
systems may be broadly divided into two classes
 depending on whether they are separable or non-separable. Systems
 for which the Hamilton-Jacobi equation is separable in a
 particular coordinate system are integrable with constants of
 motion which are typically quadratic in the respective momenta. On the
 other hand non-separable systems are typically characterized by the
 existence of
higher-order constants of motion, i.e., when the momenta are of
degree more than two \cite{CDR,GWin,Mar,MPW,MS,TW}. A recent
example being the Tremblay-Turbiner-Winternitz (TTW) system
\cite{Quesne,Rana2,TTW1,TTW2} which is directly related to the SW
potential. The so-called TTW system and
and Post-Winternitz (PW) models \cite{PW} which have recently attracted some
interest and provide concrete 
examples of superintegrable systems with non-central potentials defined on
Euclidean plane while spherical and pseudospherical generalizations of these
models represent non-isotropic superintegrable systems
on curved configuration spaces.

\noindent In the case of superintegrable systems one can identify
three possible classes \cite{CCR} namely: superseparable,
separable and nonseparable. Most of the known superintegrable
systems turn out to be superseparable, i.e., separable in more
than one coordinate system; separable superintegrable systems are
generally endowed with a mixture of quadratic and higher-order
constants of motion while for non-separable ones the constants of
motion are all of higher-order (up to the Hamiltonian).

\smallskip

While mechanical energies are obvious constants of motion for
the deduction of  additional constants a variety of methods are
usually employed. These additional integrals of motion are often
polynomials in the momenta of order higher than two. In many cases
Ra\~nada and his coworkers obtained these by the method of complex
factorization. Evans et al. \cite{Evans} and Rodriguez et al.
\cite{RTW1,RTW2} obtained them by making use
of dimensional reduction. Fordy \cite{Fordy} used the “Kaluza-Klein construction” in reverse to construct lower dimensional 
superintegrable systems from the higher dimension one. It is noteworthy to say that the 
Kaluza-Klein reduction deals with (pseudo-)Riemannian metrics, where  we consider Hamiltonians in natural form. After the reduction, the lower dimensional
Hamiltonian will have electromagnetic terms, which could turn out to be trivial. \\

Master symmetries were introduced by  Fokas and  Fuchssteiner
\cite{FF} and were also studied by  Oevel \cite{Oevel} and
Fuchssteiner \cite{Fuch}. It was first applied to nonlinear
partial differential equations (PDEs) (infinite-dimensional
Hamiltonian systems) and then to finite-dimensional systems
\cite{Dami, Fernandes}. These symmetries
are related to the existence of compatible Poisson structures and recursion operators.

Transformations mapping one integrable system to another have been
put to good use in the literature. A particular type of
transformation, known as \textit{coupling constant metamorphosis}
(CCM), was formulated by Hietarinta \textit{et al} \cite{HGDR}.
Using this technique Kress \cite{Kress} mapped the (flat space)
superintegrable system  with Hamiltonian $H = p_{x}^{2} +
p_{y}^{2} + \alpha x$ to a non-flat space superintegrable system.
It is known that all nondegenerate two-dimensional superintegrable
systems having constants quadratic in the momenta can be obtained
by coupling constant metamorphosis from those on constant
curvature space. The classification problem of classical
second-order superintegrable systems is almost settled. Most of
the results obtained rely on the use of separation of variables.
Equivalence of superintegrable systems in two dimensions are
usually studied via quadratic algebras. For a recent review
addressing the classification of second-order
superintegrable systems in two-dimensional Riemannian and
pseudo-Riemannian spaces we may cite \cite{MPW} and references
therein. It is based on the study of the quadratic algebras of the
integrals of motion and on the equivalence of different systems
under coupling constant metamorphosis.

\smallskip

Tsiganov carried out a systematic study of superintegrable Hamiltonian systems separable
in Cartesian coordinates using action-angle variables.
In a series of papers he \cite{Tsiganov1,Tsiganov2,Tsiganov3}
constructed polynomial
integrals of motion using addition theorems for the action-angle variables.
For instance, by adding action variables $I_1$ and $I_2$ one gets Hamiltonian $H$
which is in involution with the following integral of motion
$$ X = F (I_1 , I_2 , \theta_2 − \theta_1 ), \qquad \{H,F\} = 0 $$
which is functionally independent from $I_1,I_2$.
Recently Grigoriev and Tsiganov \cite{Tsiganov4} proposed the study of superintegrable systems
of Thompson's type separable in Cartesian
coordinates. In 1984 Thompson \cite{Thompson} proved superintegrability of the Hamiltonian
$$ H = p_{1}^{2} + p_{2}^{2} + a(x_1-x_2)^{-\frac{2}{2n-1}}, \qquad n \in {\Bbb Z}_+ $$. In \cite{Tsiganov4} 
Grigoriev \emph{et al} have
 shown the existence of additional integrals of motion of such superintegrable systems which are related to the famous Chebyshev theorem  \cite{Chebyshev} of binomial differentials.
Recently Gonera and Kaszubska \cite{GOK} obtained 2D superintegrable systems
defined on 2D spaces of constant curvature using actiona-angle method. These systems are separable in the
so called geodesic polar coordinates. In particular, Gonera \cite{Gonera} proved the superintegrability of the TTW model 
using action-angle methods.

\smallskip

\noindent {\bf Motivation and result} The main purpose of this
paper is to study superintegrable systems with position dependent mass
both using the method of master
symmetries due  mainly to Ra\~nada \cite{Rana1,Rana2,Rana3} and
using addition theorems for action-angle variables proposed by
Tsiganov \cite{Tsiganov1,Tsiganov2,Tsiganov3,
Tsiganov4}. We make a comparative study explore the power of these two methods to probe
superintegrable systems. We elucidate this study with various examples.

In particular, we obtain the first
integrals of the Fokas-Lagerstrom \cite{FL} and Holt \cite{Holt}
or deformed 2:1 harmonic oscillator potentials using this method.
To the best of our knowledge this is the first time
Fokas-Lagerstrom first integrals are computed via complex
factorization This factorization \cite{RRS} is obtained as a
deformation of the quadratic version of the factorization
\cite{JH,Pere} of the integrals of motion of the linear
oscillator. We show how the idea of  coupling constant
metamorphosis (CCM) can be applied to position dependent mass systems by
 finding a transformation  between a position
dependent mass 2D oscillator and Smorodinsky-Winternitz systems
and their corresponding unit mass systems. It is known that CCM in
general does not preserve the structure of the symmetry algebras,
however we can map all the conserved quantities.

\smallskip

The paper is {\bf
organized} as follows. In section 2 we recollect the method of
complex factorization  and apply it  to the Fokas-Lagerstrom and
Holt potentials. In section 3 we introduce the notion of a master
symmetry to study superintegrable systems. Finally in section 4 we
give the main result of the paper, the position dependent mass
superintegrable systems.

\section{Generalized oscillator systems}
In two dimensions an oscillatory system is typically characterized
by the Hamiltonian \be\label{OscHam} H=\frac{1}{2}(p_1^2+p_2^2)+
\frac{1}{2}\omega_0^2(n_1^2q_1^2+n_2^2q_2^2).\ee It is obvious
that such a system admits two integrals of motion given by
\be\label{B1}I_1=\frac{1}{2}(p_1^2+\omega_0^2n_1^2q_1^2),\;\;I_2=\frac{1}{2}(p_2^2+\omega_0^2n_2^2q_2^2)\ee

 A particularly simple algorithm for
finding the integrals of motion for oscillators was introduced in
\cite{Pere,JH} based on the product of powers of the complex
functions \be\label{B2}A_1=p_1+in_1\omega_0 q_1,
\;\;\;A_2=p_2+in_2\omega_0 q_2.\ee involving factorization of the
third integral. We illustrate below the procedure for finding additional
 integrals through factorization by considering the  example of a
 Fokas-Lagerstrom potential.

\bigskip

\noindent
\underline{\textbf{ Example 1: The Fokas-Lagerstrom potential}} \\
The Hamiltonian for the
 Fokas-Lagerstrom potential  is given
 by

 \be\label{FLHam}
 H=\frac{1}{2}(p_1^2+p_2^2)+\frac{1}{2}q_1^2+\frac{1}{18}q_2^2\ee
 and corresponds to the choice $n_1=1$ and $n_2=1/9$ in
 (\ref{OscHam}) with $\omega_0=1$. The Hamiltons equations of motion are
 $$\dot{q}_1=p_1,\;\;\dot{q}_2=p_2,\;\;\;\dot{p}_1=-q_1,\;\;\dot{p}_2=-\frac{1}{9}q_2.$$ The complex functions now have
 the appearance
 $$A_1=p_1+iq_1,\;\;\;\;A_2=p_2+\frac{i}{3}q_2,$$ and it follows
 that
 $$\frac{dA_1}{dt}=iA_1,\;\;\;\frac{dA_2}{dt}=\frac{i}{3}A_2.$$Consequently
 it is easily verified
 that $C=A_1A_2^{*3}$ is a complex constant of motion whose imaginary
 part
 $$C_I=p_2^2(q_1p_2-q_2p_1)+\frac{1}{27}q_2^3p_1-\frac{1}{3}q_1q_2^2p_2,$$
 yields a real constant of motion. Note the cubic dependance on the
 momenta. On the other hand the real part also gives us the constant
 of motion
 $$C_R=p_2^2(p_1p_2+q_1q_2)-\frac{q_2^2}{3}(p_1p_2+\frac{1}{9}q_1q_2).$$
As the Hamiltonian in (\ref{FLHam}) is clearly separable two
obvious integrals of motion are given by
$$I_1=\frac{1}{2}(p_1^2+q_1^2),\;\;\;I_2=\frac{1}{2}(p_2^2+\frac{1}{9}q_2^2).$$
Relabelling the constants $C_I=I_3$ and $C_R=I_4$ one may show
that they are not independent but satisfy the relation
$$I_3^2+I_4^2=16I_1I_2^3.$$

 In \cite{Rana2} the general case of a separable Hamiltonian of
 the form
 \be\label{B4} H=\frac{1}{2}(p_1^2+p_2^2)+
\frac{1}{2}\omega_0^2(n_1^2q_1^2+n_2^2q_2^2)+\frac{k_1}{2q_1^2}+\frac{k_2}{2q_2^2},\ee
was tackled in the above spirit by defining additional complex
functions
$$B_1=A_1^2+\frac{k_1}{q_1^2},\;\;\;B_2=A_2^2+\frac{k_2}{q_2^2},$$
which satisfy the equations
\be\label{B5}\frac{dB_1}{dt}=2in_1\omega_0B_1,\;\;\;\frac{dB_2}{dt}=2in_2\omega_0B_2,\ee
whence it follows that the functions
$$B_{ij}=(B_i)^{n_j}(B^*_j)^{n_i},\;\;\;i,j=1,2$$ are constants of
motion.

\bigskip

\noindent
\underline{\textbf{Example 2: Holt potential}}\\

Our next example concerns the Holt system for which the
Hamiltonian is \be\label{HoltHam}
H=\frac{1}{2}(p_1^2+p_2^2)+\frac{1}{2}(q_1^2+4q_2^2)+\frac{\delta}{q_1^2}.\ee
Defining the complex functions \be
A_1=p_1+iq_1,\;\;\;\;A_2=p_2+i2q_2,\;\;\;B_1=A_1^2+\frac{2\delta}{q_1^2}\ee
we find, using the relevant equations of motion, \be
\frac{dB_1}{dt}=2iB_1,\;\;\;\frac{dA_2}{dt}=2iA_2.\ee Consequently
it follows that $C_{12}=B_1A_2^*$ is a constant of motion with \be
Re(C_{12})=p_1^2p_2-q_1^2p_2+\frac{2\delta}{q_1^2}p_2+4q_1q_2
p_1\ee being a cubic integral of motion.

\section{Superintegrability and Master symmetries}

There exists a close relationship between superintegrability and
the concept of master symmetries. Given a Hamiltonian $H$, we say
that a function $T(q,p)$ is a generator of the constants of motion
of degree $m$ for $H$ if it satisfies the following conditions,
namely \be\label{a1}\frac{d^kT}{dt^k}\ne
0,\;\;k=1,...,m,\;\;\;\;\;\frac{d^{m+1}T}{dt^{m+1}}=0.\ee It is
evident from these conditions that $T$ is a function which
generates an integral of motion by time derivation. Following
\cite{Rana1,Rana2,Rana3} we note that for $n=2$  Hamiltonian system if $T_1$
and $T_2$ are two generators of degree $m=1$ such that $I_1$ and
$I_2$ defined by $I_k=dT_k/dt$ for $k=1,2$ are constants of motion
in involution then one can construct time-dependent constants of
motion $I_1^t$ and $I_2^t$ by means of the following definition
\be\label{a2} I_k^t=T_k-I_kt.\ee Note the linear dependence on the
time $t$ which allows us to define an additional time-independent
constant of motion given by \be\label{a3}I_{12}=T_1I_2-T_2I_1,\ee
thereby ensuring that the system is superintegrable.\\
A similar procedure exists in case of higher degree generators as
we illustrate below.

\bigskip

\noindent
\underline{\textbf{Example 3: A linear potential}}\\

We consider the Hamiltonian
given by
\be H=p_x^2+p_y^2+\alpha x. \ee
The equations of motion are given by
$$\dot{x}=2p_x,\;\;\dot{y}=2p_y,\;\;\dot{p}_x=-\alpha,\;\;\;\dot{p}_y=0$$
Two obvious first integrals are provided by
$$E_x=p_x^2+\alpha x,\;\;\;E_y=p_y$$
Let $T_1=xp_y$ then it follows that $\dot{T}_1=2p_xp_y$,
$\ddot{T}_1=-2\alpha p_y$ and $\dddot{T}_1=0$ so that
$I_1=-2\alpha p_y$ is a first integral. Similarly if we take
$T_2=yp_x$ then we find that $\dot{T}_2=2p_xp_y-\alpha y$,
$\ddot{T}_2=-4\alpha p_y$ and $\dddot{T}_2=0$ so that
$I_2=-4\alpha p_y$ is a first integral. Consequently we may deduce
the following two time-dependent first integrals using (\ref{a2})
$$I_1^t:=\dot{T}_1-I_1t=2p_y(p_x+\alpha
t),\;\;\;I_2^t:=\dot{T}_2-I_2t=2p_xp_y-\alpha y+4\alpha p_yt$$ To
deduce a time independent first integral we note that the time
derivative of
$$I_{12}=\dot{T}_1I_2-\dot{T}_2I_1$$ is by construction zero.
Hence the required time-independent first integral is
$$I_{12}=-4\alpha p_y(p_xp_y+\frac{\alpha}{2}y)$$ Since here $p_y$
is itself conserved one may scale this first integral and take,
$p_xp_y+y\alpha/2$, to be the additional (third) first integral
thus
ensuring superintegrability of the system under consideration.\\
 In fact there exists another first integral for this system  in the literature
which may be obtained as follows. Let us consider the following
$m=1$ generators namely
$$T_3=2p_xp_y-yp_x\;\;\;\;T_4=y, $$ whence it follows that
$$\dot{T}_3=2p_xp_y+\alpha y,\;\;\ddot{T}_3=0,$$
$$\dot{T}_4=2p_y,\;\;\;\ddot{T}_4=0.$$ We immediately recognize
$I_3:=2p_xp_y+\alpha y$ as representing the integral $I_{12}$
obtained earlier, so setting $I_4:=2p_y$ we obtain using
(\ref{a3}) the following first integral
$$I_{34}:=T_3I_4-T_4I_3=4\left[(xp_y-yp_x)p_y-\frac{\alpha}{4}y^2\right]$$

Relabelling these first integrals as
$$K=p_y,\;\;\; R_1=(xp_y-yp_x)p_y-\frac{\alpha}{4}y^2,\;\;\;
R_2=p_xp_y+\frac{\alpha}{2}y,\;\;\; H=E_x+E_y$$ where we have used
the notation of \cite{Kress} we note that these four integrals are
not functionally independent for they are related by
$$R_2^2+K^4-HK^2+\alpha R_1=0.$$
 Although this example is well know the fact that
 all its first integrals can be recast in the language of master
 symmetries reveals an interesting aspect of the system described by a linear potential.\\

 As our main interest is on position
 dependent mass systems we  consider below the case when the mass
 function is not a constant and study the resulting  impact on
  the methods outlined above.

\section{Position dependent mass and superintegrability}

Consider the anisotropic two dimensional harmonic oscillator with
a position dependent mass described by the following Hamiltonian:
\be\label{L1}
H=\frac{p_1^2}{2m_1(q_1)}+\frac{p_2^2}{2m_2(q_2)}+\frac{1}{2}m_1(q_1)\omega_1^2q_1^2+
\frac{1}{2}m_2(q_2)\omega_2^2q_2^2.\ee We assume that $(q_i, p_i)$
for $i=1,2$ represent a set of canonical variables and satisfy the
standard Poisson algebra $\{q_i, p_j\}=\delta_{ij}$ and
$\{q_i,q_j\}=\{p_i, p_j\}=0$. The Hamilton's equation of motion
are then given by \be\label{L2}
\dot{q}_i=\frac{p_i}{m_i(q_i)},\;\;\;\dot{p}_i=\frac{m_i^\prime(q_i)}{2m_i^2}p_i^2-\frac{1}{2}\omega_i^2
\frac{d}{dq_i}(m_i q_i^2),\;\;\;i=1,2.\ee
\subsection{Position dependent mass and complex factorization}
By analogy with (\ref{B2}) we now define the complex functions
\be\label{C2}A_i=\left[\frac{p_i}{\sqrt{m_i}}+i\omega_i\sqrt{m_i}q_i\right],\;\;\;i=1,2,\ee
it being understood that the arguments of $m_i$ are their
respective coordinates. Then
\be\label{C3}\frac{dA_i}{dt}=i\omega_i\left[1+\frac{q_i}{2}\frac{m_i^\prime(q_i)}{m_i(q_i)}\right]A_i.\ee
Defining $C_{ij}=A_iA_j^*$ it follows that
\be\label{C4}C_{ii}=\frac{p_i^2}{2m_i(q_i)}+\frac{1}{2}m_i(q_i)\omega_i^2q_i^2,\;\;\;i=1,2,\ee
are constants of motion.\\

In (\ref{C3})  suppose $\omega_i=n_i\omega_0$ while the mass
functions are such that
\be\label{C6}\left[1+\frac{q_i}{2}\frac{m_i^\prime(q_i)}{m_i(q_i)}\right]=\lambda_i\;\;\;i=1,2,\ee
where $\lambda_i$ are constants. It turns out that
$$\frac{dA_i}{dt}=i\omega_0n_i\lambda_i A_i,$$ and
\be\label{C7}C_{ij}=A_i^{n_j\lambda_j}(A_j^*)^{n_i\lambda_i},\ee
is a constant of motion. From (\ref{C6}) we recover the form of
the mass function as \be\label{C8}
m_i(q_i)=m_{0i}q_i^{2(\lambda_i-1)},\ee where $m_{0i}\;(i=1,2)$ is
a constant. As an illustration consider the case of $n_1=1$ and
$n_2=2$ while $\lambda_1=2$ and $\lambda_2=-1$ which corresponds
to the Hamiltonian
\be\label{H1}H=\frac{p_1^2}{2q_1^2}+\frac{p_2^2}{2q_2^{-4}}+\frac{1}{2}\omega_0^2q_1^4
+\frac{1}{2}\omega_0^2\frac{4}{q_2^2},\ee taking $m_{0i}=1$. These
choices result in
$$\frac{dA_1}{dt}=2i\omega_0A_1,\;\;\;\frac{dA_2}{dt}=-2i\omega_0A_2,$$
leading to the following constants of motion, viz
$$I_3 = Re(A_1A_2)=p_1p_2\frac{q_2^2}{q_1}-2\omega_0^2\frac{q_1^2}{q_2},$$
$$I_4 = Im(A_1A_2)=\left[q_1^2q_2^2p_2+\frac{2p_1}{q_1q_2}\right].$$

 Let us denote  by $H_1$ and $H_2$
 the two decoupled
 components of $H$ as appearing in (\ref{H1}).
Furthermore suppose $I_1$ and $I_2$  represent the the two
one-dimensional energies corresponding to Hamiltonians $H_1$ and
$H_2$, then an interesting property is that the Poisson bracket of
$I_1$ with $I_4$ is just $I_3$.

\subsection{Smorodinsky-Winternitz system with position dependent mass}

The $n=2$ Smorodinsky-Winternitz system has the following
Hamiltonian
\be\label{SW1}H=\frac{1}{2}(p_1^2+p_2^2)+k_0(q_1^2+q_2^2)+\frac{k_1}{q_1^2}+\frac{k_2}{q_2^2}.\ee
The generators for the particular case of $k_0=0$ was considered
in \cite{Rana1} and they are of the form
$$T_i=q_i p_i,\;\;\;i=1,2.$$
We consider below a modification of the above Hamiltonian when the
mass is position dependent (assuming  $k_0=0$), \be\label{SW2}
H=\frac{1}{2}\left(\frac{p_1^2}{m_1(q_1)}+\frac{p_2^2}{m_2(q_2)}\right)+\frac{k_1}{q_1^2}+\frac{k_2}{q_2^2}.\ee
We assume the following form of the generator of master symmetries
\be\label{SW3} T_i=f_i(q_i)p_i,\;\;\;\mbox{which satisfy the
conditions}\;\;\;\frac{dT_i}{dt}\ne 0,
\;\;\;\frac{d^2T_i}{dt^2}=0,\;\;\; i=1,2.\ee

It follows that \be\label{SW4a}\frac{dT_i}{dt}=X_i(q_i)p_i^2+
2k_i\frac{f_i}{q_i^3}\ne 0,\ee
\be\label{SW4b}\frac{d^2T_i}{dt^2}=\left[X_i^\prime(q_i)+\frac{m_i^\prime}{m_i}X_i(q_i)\right]\frac{p_i^3}{m_i}
+\left[X_i(q_i)\frac{4k_i}{q_i^3}+\frac{2k_i}{m_i}\left(\frac{f_i}{q_i^3}\right)^\prime\right]p_i=0\ee
where
$$X_i(q_i)=\left(\frac{f_i^\prime}{m_i}+f_i\frac{m_i^\prime}{2m_i^2}\right)\;\;\;i=1,2.$$
and the $^\prime$ denotes differentiation with respect to the
appropriate argument. Equating the coefficients of the powers of
$p_i$ from (\ref{SW4b}) we obtain the following equations for
determining the function $f_i$ and the  mass $m_i$,
\be\label{SW5a}X_i^\prime(q_i)+\frac{m_i^\prime}{m_i}X_i(q_i)=0\ee
\be\label{SW5b}X_i(q_i)\frac{2}{q_i^3}+\frac{1}{m_i}\left(\frac{f_i}{q_i^3}\right)^\prime=0.\ee

Eqn (\ref{SW5a}) implies
\be\label{SW6}m_iX_i(q_i)=\lambda_i,\;\;\;i=1,2\ee with
$\lambda_i$ being a constant. On the other hand from (\ref{SW5b})
upon using (\ref{SW6}) we find  that
\be\label{SW7}f_i(q_i)=\lambda_iq_i+\mu_iq_i^3,\;\;\;i=1,2\ee
where $\mu_i$ is also a constant. The explicit form of the mass
function now follows from (\ref{SW6}) and is given by
\be\label{SW8} m_i(q_i)=\frac{1}{(\lambda_i+\mu_iq_i^2)^3},
\;\;\;i=1,2.\ee

 Thus it is evident that when
 $\mu_i=0$ the mass function  becomes a constant while
$f_i=q_i$. This precisely corresponds to the situation considered
in \cite{Rana1}. In our case the time-independent first integrals,
 given by $dT_i/dt$, are \be\label{SW9}
I_i=(\lambda_i+\mu_iq_i^2)\left[\lambda_i(\lambda_i+\mu_iq_i^2)^2p_i^2+\frac{2k_i}{q_i^2}\right],\;\;i=1,2.\ee

The corresponding time-dependent functions therefore have the form
\be\label{SW10}I_i^t=(\lambda_iq_i+\mu_iq_i^3)p_i-
\left[(\lambda_i+\mu_iq_i^2)\left(\lambda_i(\lambda_i+\mu_iq_i^2)^2p_i^2+\frac{2k_i}{q_i^2}\right)\right]t,\;\;i=1,2\ee

Thus from (\ref{a3}) it at once follows that there exists a time
-independent constant of motion given by \be\label{SW11}
I_{12}=(\lambda_1+\mu_1q_1^2)(\lambda_2+\mu_2q_2^2)
\left[\left((\lambda_2+\mu_2q_2^2)^2\lambda_2p_2^2+\frac{2k_2}{q_2^2}\right)q_1p_1
-\left((\lambda_1+\mu_1q_1^2)^2\lambda_1p_1^2+\frac{2k_1}{q_1^2}\right)q_2p_2\right].\ee

In a similar manner it may be shown that  for the Hamiltonian
\be\label{x1}H=\frac{p_x^2}{2m_x}+\frac{p_y^2}{2m_y}+k_2x+\frac{k_3}{y^2}\ee
where $m_x$ and $m_y$ are functions of $x$ and $y$ respectively
 the generators of the first integrals are given by
$$T_1=(B_1-C_1x)p_x,\;\;\;\;m_x(x)=\frac{A_1}{(B_1-C_1x)^3}$$
$$T_2=(B_2y+C_2y^3)p_y,\;\;\;\;\;m_y(y)=\frac{A_2}{(B_2+C_2y^2)^3}$$
with a time-independent first integral
$$I_1=\frac{C_1}{2A_1}(B_1-C_1x)^3p_x^2-k_2(B_1-C_1x)$$

\section{Metamorphosis and duality between position dependent mass systems and unit mass
oscillators}

In this section we show that a duality transformation exist
between position dependent mass $2D$ oscillator and constant mass
2D oscillator. In the 1980's, a number of papers were devoted to the
investigation of certain duality properties of pairs of Hamiltonians.
The underlying idea was based on the work of  Hietarinta \emph{et al}.
\cite{HGDR} and received a lot of attention as the result of one integrable
system automatically implied the existence of another type of (or
version) integrable system.

\smallskip

In order to illustrate this feature we consider the standard harmonic oscillator, whose Lagrangian and Hamiltonian are given by
$L = \frac{1}{2}v_{x}^{2} - \frac{1}{2}\omega^2x^2$ and
$ H = \frac{1}{2}p_{x}^{2} + \frac{1}{2}\omega^2x^2$, respectively.
Consider the following change of variables
$$
(x,v_x)  = (q,v_q); \qquad  x = q^{\lambda}, \,\,\,\,\, v_x = \lambda q^{\lambda -1}v_q.
$$
The transformed Lagrangian and Hamiltonian in new coordinates are given by
$$
{\tilde L} = \frac{1}{2}\lambda^2 q^{2(\lambda -1)}v_{q}^{2} -  \frac{1}{2}\omega^2q^{2\lambda}, \qquad
 {\tilde H} = \frac{1}{2}\frac{1}{\lambda^2} \frac{p_{q}^{2}}{q^{2(\lambda -1)}} +
\frac{1}{2}\omega^2q^{2\lambda}.
$$
Let us introduce the following notation \be m(q) = q^{2(\lambda
-1)}, \qquad \omega = n \omega_0, \ee then  ${\tilde H}$ can be
written as follows \be {\tilde H} = \frac{1}{\lambda^2} [
\frac{1}{2}\frac{p_{q}^{2}}{m(q)} +
\frac{1}{2}(\lambda^2n^2)\omega_{0}^{2} m(q)q^2 ], \ee which can
be normalized $ {\tilde H} = \frac{1}{\lambda^2}{\tilde
H}_{\lambda}$. A nonlinear oscillator with a position dependent
mass $m(q)$ is the
 {\it standard harmonic oscillator} but just written in a new system of coordinates.

\smallskip

Next we study the inverse transformation.
 Under the
 transformation
 \be\label{C2} Q_1=\frac{1}{2}q_1^2,\;\;\;
 P_1=\frac{p_1}{q_1},\;\;\;
 Q_2=-\frac{1}{q_2},\;\;\; P_2=p_2q_2^2,\ee the Hamiltonian (\ref{H1})
 reduces to
 \be\label{H2}\bar{H}=\frac{1}{2}P_1^2+\frac{1}{2}(2\omega_0)^2Q_1^2+\frac{1}{2}P_2^2+\frac{1}{2}(2\omega_0)^2Q_2^2.\ee
 This clearly corresponds to the constant mass scenario. The
 corresponding first integrals $I_3$ and $I_4$ now become  just
 the Fradkin tensor and the angular momentum respectively, i.e.,
 $$I_3=P_1P_2+(2\omega_0)^2Q_1Q_2,\;\;\; I_4=2(Q_1P_2-Q_2P_1).$$

\subsection{Duality of PDM Smorodinsky-Winternitz equation}

In order to extend the above idea to the PDM scenario
let us consider the  Hamiltonian of the position dependent mass SW equation
\be
H = \frac{1}{2}p_{1}^{2}(1 + q_{1}^{2})^3 +   \frac{1}{2}p_{2}^{2}(1 + q_{2}^{2})^3 + \frac{k_1}{q_{1}^{2}}
+ \frac{k_2}{q_{2}^{2}}.
\ee
We define $P_1 = p_1(1 + q_{1}^2)^{3/2}$ and $P_2 = p_2(1 + q_{2}^2)^{3/2}$, and fix the form of  $Q_1 = Q_1(q_1,p_1)$
 via the canonical requirement
$$
\{Q_1,P_1\} = \frac{\partial Q_1}{\partial q_1}\frac{\partial P_1}{\partial p_1} -
\frac{\partial Q_1}{\partial p_1}\frac{\partial P_1}{\partial q_1} = 1,
$$
from which we obtain
$$
Q_{1q_1}(1 + q_{1}^{2}) - 3p_pq_1(1 + q_{1}^{2})^{1/2}Q_{1p_1} = (1 + q_{1}^{2})^{- 1/2}.
$$
The corresponding Lagrange system of equations is
$$
\frac{dq_1}{(1 + q_{1}^{2})} = \frac{dp_1}{-3p_1q_{1}} = \frac{dQ_1}{(1 + q_{1}^{- 1/2})}.
$$
and yields the characteristics
$$C_1=p_1(1+q_1^2)^{3/2},\;\;\;C_2=Q_1-\frac{q_1}{\sqrt{1+q_1^2}}$$ and hence the general solution $C_2=F(C_1)$. Choosing the arbitrary function $F$ to be the null function immediately yields a particularly simple solution for $Q_1$ namely
 $ Q_1 = \frac{q_1}{\sqrt{1 + q_{1}^{2}}}$ and similarly
$ Q_2 = \frac{q_2}{\sqrt{1 + q_{2}^{2}}}$.

\begin{prop}
Let $ H =  \frac{1}{2}p_{1}^{2}(1 + q_{1}^{2})^3 +   \frac{1}{2}p_{2}^{2}(1 + q_{2}^{2})^3 + \frac{k_1}{q_{1}^{2}}
+ \frac{k_2}{q_{2}^{2}}$ be the position dependent Smorodinsky-Winternitz system, then
under the transformation $Q_i = \frac{q_i}{\sqrt{1 + q_{i}^{2}}}$ and $P_i = p_i(1 + q_{i}^2)^{3/2}$
$H$ becomes the Hamiltonian of the unit mass SW equation upto an additive constant viz
$$H\longrightarrow \tilde{H}=\frac{1}{2}P_{1}^{2}+\frac{1}{2}P_{2}^{2}+\frac{k_1}{Q_1^2}+\frac{k_2}{Q_2^2}+(k_1+k_2)$$
and the transformation may be used to map
the integrals of motion of the two systems.
\end{prop}

\section{Generalized oscillatory systems and master symmetries}
Let us consider once again the Hamiltonian of (\ref{L1}) and the
associated equations of motion as given by (\ref{L2}). Following
Ranada \cite{Rana1} we define the variable \be\label{L3}
u_i=\frac{\omega_if_i(q_i)p_i}{E_i},\;\;\;E_i=\frac{p_i^2}{2m_i(q_i)}+\frac{1}{2}m_i(q_i)\omega_i^2q_i^2,\;\;\;i=1,2\ee
where it is easy to verify,  in view of (\ref{L2}) that
$dE_i/dt=0$. Suppose $T_i=\arcsin(u_i)$ so that \be \frac{dT_i}{dt}=\frac{1}{\sqrt{1-u_i^2}}\frac{du_i}{dt}.\ee
Employing the definitions given in (\ref{L3}) it follows that
$$\frac{dT_i}{dt}=\frac{2\omega_i}{\sqrt{E_i^2-\omega_i^2f_i^2p_i^2}}
\left[f_i\left(\frac{m_i^\prime(q_i)}{2m_i^2}p_i^2-\frac{1}{2}\omega_i^2
\frac{d}{dq_i}(m_i
q_i^2)\right)+\frac{f_i^\prime}{m_i}p_i^2\right].$$

Next let us demand that $dT_i/dt=2\omega_i\lambda_i$ where $\lambda_i$
is an arbitrary constant, which essentially means that
\be\label{L4}\left[f_i\left(\frac{m_i^\prime(q_i)}{2m_i^2}p_i^2-\frac{1}{2}\omega_i^2
\frac{d}{dq_i}(m_i
q_i^2)\right)+\frac{f_i^\prime}{m_i}p_i^2\right]=\lambda_i\sqrt{E_i^2-\omega_i^2f_i^2p_i^2}.\ee

It follows  that $d^2T_i/dt^2=0$ and hence $T_i$ is a
generator of a master symmetry for the PDM Hamiltonian (\ref{L1}).
Upon squaring both sides of (\ref{L4}) and equating the
coefficients of the different powers of $p_i$ we obtain the
following set of equations $(i=1,2)$, namely
\begin{align}f_i\left(\frac{m_i^\prime}{2m_i}+\frac{f_i^\prime}{f_i}\right)
&=\pm\lambda_i,\\
q_i^2f_i\left[f_i\left(\frac{m_i^\prime}{2m_i}+\frac{f_i^\prime}{f_i}\right)\right]\frac{(m_iq_i^2)^\prime}{(m_iq_i^2)}
&=2\lambda_i^2(2f_i^2-q_i^2),\\
\frac{(m_iq_i^2)^\prime}{(m_iq_i^2)}&=\pm\frac{2\lambda_i}{f_i}.\end{align}
By eliminating the terms involving the mass $m_i$ it readily
follows that the only acceptable form of the unknown function
$f_i$ is given by \be\label{L6} f_i(q_i)=q_i, \;\;\;i=1,2.\ee This
is precisely the form with which the author of \cite{Rana1} began.
However the interesting feature here is that in presence of a
position dependent mass term with this form of the function $f_i$
one finds from the above set of equations that the mass function
has either of the following two forms depending on the $\pm$ sign,
viz
\begin{align} m_i(q_i) &=q_i^{2(\lambda_i-1)},\;\;\;\mbox{for +
sign}\\
m_i(q_i) &=q_i^{-2(\lambda_i+1)},\;\;\;\mbox{for -
sign}\end{align} The specific choice $\lambda_i=1$ leads to the
case where each $m_i=1$ and was considered in \cite{Rana1}. Even
with this choice of $\lambda_i$ there exists a second possibility
(corresponding to the negative sign) wherein $m_i=q_i^{-4}$ for
which the Hamiltonian of (\ref{L1}) assumes the following form
\be\label{L7}H=\sum_{i=1}^2\left[\frac{1}{2}p_i^2q_i^4+\frac{1}{2}\frac{\omega_i^2}{q_i^2}\right]\ee
The equation of motion for $q_i$  following from the above
Hamiltonian represent second-order ordinary differential equations
of the Li\'{e}nard -II type namely
\be\label{L8}\ddot{q_i}+\frac{2}{q_i}\dot{q}_i^2-\omega_i^2q_i=0,\;\;\;i=1,2.\ee

\section{Action-angle method: Tsiganov's approach}
Recently the concept of action-angle variables has been employed to derive an additional first integral typically
for the systems  considered in this paper. The method has proved to be complimentary to that of Master symmetries and is 
in some sense closer in spirit to the very notion of integrability itself being dependent on
 action-angle variables. To this end we consider  example 3 again where
$$H=p_1^2+p_2^2+\alpha q_1$$
for which two first integrals are obvious, namely
$$I_1=p_1^2+\alpha q_1, \;\;\; I_2=p_2^2$$
We define the angle variables
$$\phi_1=\frac{\partial}{\partial I_1}\int p_1 dq_1=\frac{\partial}{\partial I_1}\int^{q_1}\sqrt{I_1-\alpha x}dx
=\frac{1}{2}\int^{q_1}\frac{dx}{\sqrt{I_1-\alpha x}}$$ and similarly
$$\phi_2=\frac{1}{2}\int^{q_2}\frac{dx}{\sqrt{I_2}}$$
One can easily verify that
$$\{\phi_1, I_1\}=\{\phi_2, I_2\}=1, \;\;\{\phi_1, \phi_2\}=\{I_1, I_2\}=0$$
Following \cite{Tsiganov4} \emph{et al} we note that any function $F=F(I_1, I_2, \phi_1-\phi_2)$ can be regarded as an additional functionally independent first integral. We verify this by taking $X=\phi_1-\phi_2$. It is possible to evaluate $X$ explicitly to obtain
$$X=\phi_1-\phi_2=\frac{1}{2}\left[\int^{q_1}\frac{dx}{\sqrt{I_1-\alpha x}}-\int^{q_2}\frac{dx}{\sqrt{I_2}}\right]
=-\frac{1}{\alpha}\sqrt{I_1-\alpha q_1}-\frac{q_2}{2\sqrt{I_2}}$$
The equations of motion following from the above Hamiltonian are $$\dot{q_1}=p_1,\;\; \dot{p_1}=-\alpha,\;\; \dot{q}_2=p_2,\;\; \dot{p}_2=0$$ and one can verify that
$$\frac{dX}{dt}=\dot{\phi}_1-\dot{\phi}_2=0$$ where
$$X=-\frac{1}{\alpha}p_2^{-1}(p_1p_2+\frac{\alpha}{2}q_2)=-\frac{1}{\alpha \sqrt{I_2}}(p_1p_2+\frac{\alpha}{2}q_2),$$
which is clearly consistent with the earlier result.

As a second illustration we consider the following example taken from Ranada \cite{Rana4}
\be\label{RR1}H=\frac{1}{2}(p_1^2+p_2^2)+\frac{1}{2}\omega_0^2(n_1q_1^2+n_2^2q_2^2)+\frac{k_1}{2q_1^2}+k_2q_2.\ee
Once again it is obvious that
\be\label{RR2}I_1=\frac{1}{2}p_1^2+\frac{1}{2}\omega_0^2n_1^2q_1^2+\frac{k_1}{2q_1^2},\;\;\;I_2=\frac{1}{2}p_2^2+
\frac{1}{2}\omega_0^2n_2^2q_2^2+k_2q_2,\ee
are first integrals. Proceeding as in the previous example we define
the coresponding angle variables
\be\label{RR3}\phi_1=\frac{\partial}{\partial I_1}\int^x p_1 dx^\prime=\frac{1}{\sqrt{2}}\int^x
\frac{dx^\prime}{\sqrt{I_1-\frac{1}{2}\omega_0^2n_1^2x^{\prime 2}-\frac{k_1}{2x^{\prime 2}}}},\ee
\be\label{RR4}\phi_2=\frac{\partial}{\partial I_2}\int^y p_2 dy^\prime=\frac{1}{\sqrt{2}}\int^y
\frac{dy^\prime}{\sqrt{I_1-\frac{1}{2}\omega_0^2n_2^2y^{\prime 2}-k_2y^\prime}}.\ee

\smallskip

According to the Chebyshev theorem \cite{Chebyshev}  on integrals of  differential binomials of the form
$$ \int x^m(a + bx^n)^p \,dx ,$$ such integrals can be evaluated in terms of elementary functions if and only if\\
(a) $p$ is an integer, then we expand $(a + bx^n )^p$ by the binomial formula in order to rewrite the
integrand as a rational function of simple radicals $x^{j/k}$. By a simple substitution $x=t^r$ we
remove the radicals entirely and obtain integral on
rational function.\\
(b)$m+1/n$ is an integer, then setting $t = a + bx^n$ we convert the integral to $\int t^p (t-a)^{m=1/n - 1}dt$. \\
(c)$m+1/n + p$ is an integer, then we transform the integral by factoring out $x^n$ and resultant
new integral of the differential binomial belongs to case (b).

\smallskip
In view of the above 
one may evaluate these integrals in (\ref{RR3}) and (\ref{RR4}) explicitly and obtain their difference as
\be\label{RR5}\phi_1-\phi_2=\frac{1}{2\omega_0n_1}
\arcsin\left(\frac{\frac{\omega_0n_1}{\sqrt{2}}\left(x^2-\frac{I_1}{\omega_0^2n_1^2}\right)}
{\sqrt{\frac{I_1^2}{2\omega_0^2n_1^2}-\frac{k_1}{2}}}\right)
-\frac{1}{\omega_0n_2}\arcsin\left(\frac{\frac{\omega_0n_2}{\sqrt{2}}\left(y+\frac{k_2}{\omega_0^2n_2^2}\right)}
{\sqrt{I_2+\frac{k_2^2}{2\omega_0^2n_2^2}}}\right).\ee
It is possible to verify, using the Hamiltons equations of motion, that $X:=\phi_1-\phi_2$ is a constant of motion.
Moreover, using the complex logarithmic version of the $\arcsin(z)$ function, namely
$$\arcsin(z)=-i\log(iz+\sqrt{1-z^2})$$ we may exponentiate the constant of motion to obtain
\be\label{RR6}e^{i2n_1n_2\omega_0X}=(iz_1+\sqrt{1-z_1^2})^{n_2}(iz_2+\sqrt{1-z_2^2})^{-2n_1}\ee where
$$z_1=\left(\frac{\frac{\omega_0n_1}{\sqrt{2}}\left(x^2-\frac{I_1}{\omega_0^2n_1^2}\right)}{
\sqrt{\frac{I_1^2}{2\omega_0^2n_1^2}-\frac{k_1}{2}}}\right),\;\;\;
z_2=\left(\frac{\frac{\omega_0n_2}{\sqrt{2}}\left(y+\frac{k_2}{\omega_0^2n_2^2}\right)}
{\sqrt{I_2+\frac{k_2^2}{2\omega_0^2n_2^2}}}\right).$$
Next we look at the Smorodinsky-Winternitz system for which the Hamiltonian is given by
$$H=\frac{1}{2}p_1^2+k_0q_1^2+\frac{k_1}{q_1^2}+\frac{1}{2}p_2^2+k_0^\prime q_1^2+\frac{k_1^\prime}{q_1^2}$$
As usual two obvious first integrals are
$$I_1=\frac{1}{2}p_1^2+k_0q_1^2+\frac{k_1}{q_1^2}, \;\;I_2=\frac{1}{2}p_2^2+k_0^\prime q_2^2+\frac{k_1^\prime}{q_2^2}$$
A similar calculation as in the previous case gives the additional first integral
\be\label{RR7}X_{SW}=\frac{1}{2\sqrt{2}}\left[\frac{1}{\sqrt{k_0}}
\arcsin\left(\frac{q_1^2-\frac{I_1}{2k_0}}{\frac{C_1}{\sqrt{k_0}}}\right)-
\frac{1}{\sqrt{k_0^\prime}}
\arcsin\left(\frac{q_2^2-\frac{I_2}{2k_0^\prime}}{\frac{C_2}{\sqrt{k_0^\prime}}}\right)\right],\ee
where $$C_1^2=\frac{I_1^2}{4k_0}-k_1, \;\;\;C_2^2=\frac{I_2^2}{4k_0^\prime}-k_1^\prime$$
Similar manipulations, using the complex logarithmic form of the inverse sine function,  now lead to the integral (complex)

\be\label{RR8}I_{SW}=\left[\frac{\sqrt{2k_0}p_1q_1-i\left(\frac{p_1^2}{2}-q_1^2+\frac{k_1}{q_1^2}\right)}
{\sqrt{I_1^2k_0-4k_0k_1}}\right]^{\sqrt{k_0^\prime}}
\left[\frac{\sqrt{2k_0^\prime}p_2q_2-i\left(\frac{p_2^2}{2}-q_2^2+\frac{k_1^\prime}{q_2^2}\right)}
{\sqrt{I_2^2k_0^\prime-4k_0^\prime k_1^\prime}}\right]^{-\sqrt{k_0}}.\ee

\section{Final comments}

It is well known that there exist a variety of methods for  proving the  superintegrability of various potentials 
 each having its own merits as far as computational flexibility is concerned. In this paper we have considered two approaches namely the  master symmetry procedure and the method based on action-angle variables. The latter requires the explicit evaluation of certain integrals which has been aided  by the Chebyshev theorem on binomial differentials. On the other hand the former relies more on the existence of a series of constructs leading ultimately to an integral of motion by time derivation. In this context we have also looked at position dependent mass versions of some of the existing superintegrable systems.

 We have illustrated and compared the two methods considered in this paper with several examples such as the Fokas-Lagerstorm,
Holt type potential, Smorodinsky-Winternitz type equation and have demonstrated how in many cases the action-angle method, which plays a fundamental role in classical and quantum mechanics, captures the phenomena of the master symmetry method.
It is known that the classical action-angle variables are defined only in some domain of the phase
space which in many cases overlaps with domain of complex factorization, which is
the heart of master symmetry method. In a sense the two almost equivalent methods have
the  advantage of possessing a great degree of elegance and
simplicity which reflects their inherent robustness. 

\smallskip
 Finally it will obviously be interesting to study the quantum counterpart of the superintegrable Hamiltonian
using quantum analogs of the action-angle variables, which  play an important role in
semi-classical quantization and also to explore quantum analogs using master symmetry
or complex factorization method.

\section*{ Acknowledgement}
We would like to express our gratitude to Andrey Tsiganov for his careful reading and valuable comments.
We are also thankful to Manuel Ranada and Pepin Carinena
for their valuable inputs and enlightening discussions on
superintegrable systems.
PG gratefully acknowledge support from
Professor G. Rangarajan and IISc Mathematics Department where a
part of the work was begun.

\end{document}